\documentclass[aps,prb,showpacs,reprint]{revtex4-1}

\usepackage{graphicx}
\usepackage{amsmath}
\usepackage{bbm}

\begin{document}
\title
{Effect of the inter-subband spin-orbit interaction on the spin transistor operation}
\author{P. W\'ojcik}
\email[Electronic address: ]{pawel.wojcik@fis.agh.edu.pl}
\author{J. Adamowski}
\affiliation{AGH University of Science and Technology, Faculty of
Physics and Applied Computer Science, al. Mickiewicza 30,
Krak\'ow, Poland}

\begin{abstract}
We consider the electron transport in the Datta-Das spin transistor within the two-subband model
taking into account the intra- and inter-subband spin-orbit (SO) interaction and
study the influence of the inter-subband SO coupling on the spin-transistor operation. Starting
from the model, in which the SO coupling constants are treated as parameters, we show
that the inter-subband SO interaction strongly affects the ordinary conductance oscillations
predicted for the transistor with the single occupancy. Interestingly, we find that even  in  the
absence of the intra-subband SO interaction, the
conductance oscillates as a function of the inter-subband SO coupling constant. This phenomenon is
explained as resulting from the inter-subband transition with spin-flip. Next, we consider
the realistic spin transistor model based on the gated
Al$_{0.48}$In$_{0.52}$As/Ga$_{0.47}$In$_{0.53}$As
double quantum well, for which the SO coupling constants are determined by the
Schr\"{o}dinger-Poisson approach. We show that the SO coupling constants rapidly change
around $V_g=0$, which is desirable for the spin transistor operation. We demonstrate that for high
electron densities the inter-subband SO interaction starts to play the dominant role. The strong
evidence of this interaction is the reduction of the conductance for gate voltage  $V_g=0$, which
leads to the reduction of the on/off conductance ratio.

\end{abstract}

\maketitle
\section{Introduction}
The coherent manipulation of the electron spin in semiconductor materials via the coupling of the
electron's motion with its spin degree of freedom is a key ingredient in most spintronic
devices.~\cite{Szumniak2012} The special place among them belongs to the spin field effect
transistor (spin-FET)~\cite{Datta1990} in which the electrically tunable
spin-orbit interaction of Rashba~\cite{Rashba1984} is used to control - via the spin rotation - the
electric current between ferromagnetic source and drain. However, the experimental
realization of the functional spin-FET encounters serious physical obstacles, i.e. the low
efficiency of the spin injection from ferromagnet into
semiconductor due to the resistance mismatch~\cite{Schmidt2000} and the spin
relaxation induced mostly by the Dyakonov-Perel mechanism.~\cite{Fabian2007} Both these effects
lead to the low electrical signal and the low ratio of the "on-conductance" to the "off-conductance"
in the first experimental realization of the spin-FET.~\cite{Koo2009,Wunderlich2010} Note, that the
maximum value of the on/off conductance ratio defined as $G_{on}/G_{off}=(1+P_SP_D)/(1-P_SP_D)$,
where $P_S(P_D)$ is the spin injection (detection) efficiency in the source (drain), has a value
$2.92$ for $P_S=P_D=70 \%$ (the highest spin injection efficiency reported at room
temperature\cite{Salis2005}), which is insufficient for the electric circuit application.
Therefore, the basic condition which has to be meet in the experimental setup of the spin-FET is the
spin injection (detection) nearly equals to $100 \%$ -  the ratio $G_{on}/G_{off}=10^5$, adequate
for the modern electronics, requires $P_S=P_D=99.9995 \%$.~\cite{Spintronics} This requirement 
can be satisfied only by the use of the semiconductor spin filters such as magnetic
resonant tunneling diodes~\cite{Wojcik2012,Wojcik2013} or quantum point contacts
(QPC) with the lateral Rashba SO interaction.~\cite{Debray2009,Khoda2012,Wan2009,Nowak2013} The
latter have been successfully used as the
spin injector and detector in the recent experiment,\cite{Chuang2015, Alomar2016} in which about
$10^5$ times greater conductance oscillations have been observed as compared to the conventional
spin-FET based on ferromagnets.~\cite{Koo2009} The further improvement of the spin transistor
operation involves the suppression of the spin relaxation. For this purpose the layer conduction
channel with two dimensional electron gas (2DEG) should be replaced
by the nanowire~\cite{Wojcik2014} in which the Dyakonov-Perel mechanism of the spin relaxation is
strongly suppressed by the motional narrowing effect.\cite{Holleitner2006,Kwon2007}
Another concept assumes equating the Rashba and Dresselhaus term~\cite{Schliemann2003, Koralek2009}
which generates the persistent spin helix state with extraordinary long spin lifetime. Nevertheless,
this concepts~\cite{Schliemann2003,Kunihashi2012,Yoshizumi2016} of the spin transistor is still
waiting for the experimental realization. 

The alternative spin transistor design in which
the spin signal is observed over the distance 50~$\mu$m has been recently demonstrated by
Betthausen et al. in Ref.~\onlinecite{Betthausen2012}.
In this design, the spin transistor action is generated by the Landau-Zener transitions, which
occur in the the combined homogeneous and helical magnetic fields. The latter is
generated by the ferromagnetic stripes located above the conduction channel made of the magnetic
semiconductor. As shown in Refs.~\onlinecite{Betthausen2012, Saarikoski2014}, by
keeping the transport in the adiabatic regime, the spin state is protected
against the electron scattering on defects. The switching into the non-adiabatic regime generates
the additional conductance dips, which result from the resonant Landau-Zener
transitions.~\cite{Wojcik2015_SST} Although the alternative spin-FET~\cite{Betthausen2012} seems to
be characterized by the long spin lifetime, it requires the application of
the external homogeneous magnetic field, which is difficult to be applied in the integrated
circuit. For this reason, in our recent paper\cite{Wojcik2016} we have proposed analogous design, in
which the spin transistor action is generated by all-electric means with the use of the lateral
Rashba SO interaction.

Most of the theoretical studies and experimental realizations of the spin transistor reported
so far have been based on 2DEG fabricated in the narrow n-type AlInAs/GaInAs well.\cite{Koo2009,
Wunderlich2010} In the sufficiently narrow quantum well the electrons occupy only the first subband,
i.e. we are dealing with the lowest-energy state occupancy. However, 
the recent interest of researchers is directed towards the systems with the wide and coupled
quantum wells\cite{Bentmann2012,Hernandez2013,Hu1999} with double occupancy (two lowest-energy
subbands are occupied), which leads to interesting physical effects such
as band anticrossings or spin mixing. The SO interaction in 2DEG
quantum well with two subbands  has been studied by Bernardes et al. in
Ref.~\onlinecite{Bernardes2007}. The inter-subband-induced SO interaction has been found which
results from coupling between states with opposite parity. This inter-subband SO interaction,
quadratic in the momentum, can give raise to
interesting physical phenomena, e.g. unusual Zitterbewegung\cite{Bernardes2007} or intrinsic spin
Hall effect in symmetric quantum wells.\cite{Hernandez2013,Khaetskii2016} All these new
phenomena motivated us to
investigate the spin-FET based on the conduction channel with double
occupancy and analyze the influence of the inter-subband-induced SO interaction on the spin
transistor operation.
\begin{figure*}[ht]
\begin{center}
\includegraphics[scale=0.6]{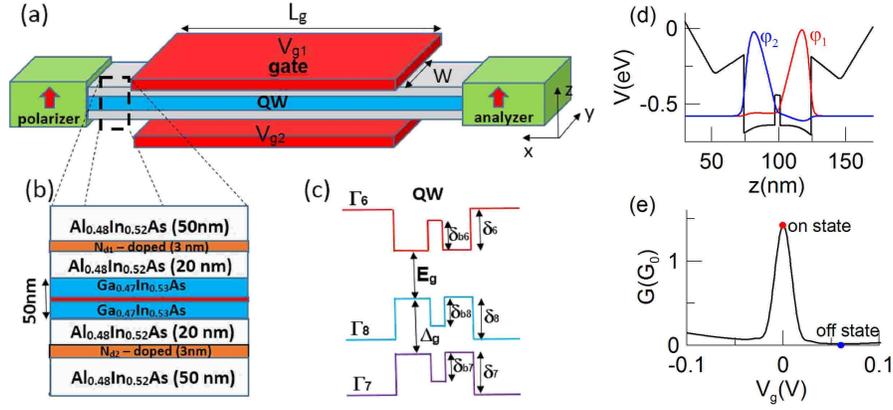}
\caption{(a) Schematic of the spin transistor. Nanowire of width $W$ is located between two
leads acting as the spin polarizer and analyzer. The spin dynamics in the conduction channel is
controlled by the voltages applied to the gates $V_{g1}$ and $V_{g2}$.  (b) Cross section
of Al$_{0.48}$In$_{0.52}$As/Ga$_{0.47}$In$_{0.53}$As double quantum well with a central
barrier. (c) Band profile for the double quantum well with the central barrier.
(d) Self-consistent potential energy profile and the corresponding wave functions $\varphi_1$ and
$\varphi _2$. (e)  $G(V_g)$ characteristics of the spin transistor with the marked on and off
state.}
\label{fig1}
\end{center}
\end{figure*}

In the present paper, we consider the electron transport in the Datta and Das spin transistor
architecture within the two subband model, which allows us to include the intra- and inter-subband
SO couplings. Starting from the model, in which the values of the SO coupling constants are
treated as the parameters, we analyze the influence of the inter-subband-induced SO interaction on
the conductance and answer the question how this type of SO interaction affects the operation of
spin transistor. Next, we consider the realistic double quantum well with the applied external
gate voltages for different electron concentrations. Following the method proposed in
Ref.~\onlinecite{Calsaverini2008}, based on the $8\times 8$ Kane model within the $\mathbf{k}\cdot
\mathbf{p}$ approximation, we determine the intra- and inter-subband induced SO coupling constants 
via the self-consistent Schr\"{o}dinger-Poisson procedure. These values are used in the conductance
calculations performed by the scattering matrix method. We reproduce the resonant
behavior of the SO coupling constants reported for a double quantum well.\cite{Calsaverini2008} This
resonant behavior for which the values of the SO parameters change abruptly near the zero gate
voltage is suitably for the spin transistor application in which the on/off transition should be
realized in the narrow voltage range. By calculating the conductance for different gate voltages, we
analyze the spin transistor operation for different electron concentrations $n_e$ and find that for
high $n_e$ the inter-subband-induced SO interaction start to play a crucial role leading to the
suppression of the on/off conductance ratio. Finally, the spin transistor operation is analyzed in
the context of the coupling between the quantum wells which is determined by the width
of the central barrier.

The paper is organized as follows: in section ~\ref{sec2} we introduce the model of the
nanostructure and briefly review the Kane Hamiltonian, which leads to the formulas for the intra-
and inter-subband SO coupling constants. Next, we describe the self-consistent
Schr\"{o}dinger-Poisson method used to the SO coupling constants calculations. Finally, we derive
the $4 \times 4$ Hamiltonian in two subband model used to the electronic transport calculations
within the scattering matrix approach.
In section~\ref{sec3} we present our results starting from these obtained for the model
in which the values of the SO coupling constants are treated as the parameters and going to
the realistic double quantum well heterostructure. The summary is contained in sec.~\ref{sec4}.

\section{Theoretical model}
\label{sec2}
\subsection{Model of nanostructure}
\label{sec2a}
We consider the Datta and Das spin transistor architecture. Accordingly, the
nanowire of width $W$ is located between two reflectionless leads acting as the
spin polarizer and analyzer [see Fig.~\ref{fig1}(a)]. In order to ensure the high value of the
on/off conductance ratio we assume 100\% spin injection (detection) efficiency of the
contacts, which as shown by recent experiments,\cite{Debray2009,Khoda2012,Wan2009} can be achieved
using the QPC with the lateral Rashba SO interaction.
Figure~\ref{fig1}(b) presents the cross-section of the layer heterostructure in the grown direction.
We consider the Al$_{0.48}$In$_{0.52}$As/Ga$_{0.47}$In$_{0.53}$As
double quantum well (width $50$ nm) with a central barrier Al$_{0.3}$In$_{0.7}$As with width $w_b$
which determines the coupling between the conduction electron states in the quantum wells. The
nanostructure contains two $n-$doped layers with donor concentrations $N_{d}=4\times 10
^{18}$~cm$^{-3}$ and width $3$~nm located on either side of the quantum well, $20$ nm away from the
well interface. In this nanodevice, the Rashba SO interaction can be tunned by the
external gates  with the lengths $L_g$ located below and above the quantum well, $50$ nm away from
the doping layers. By applying the suitably chosen voltages to these gates the spin transistor can
be electrically switched between the on to off states as shown in Fig.~\ref{fig1}(e).

\subsection{Hamiltonian with SO interaction}
Here we briefly present the derivation of an effective Hamiltonian for conduction electrons with SO
interaction. We start from the $8 \times 8$ Kane Hamiltonian
for the layer heterostucture, which in the block form is given
by\cite{Calsaverini2008,Fabian2007} 
\begin{equation}
\label{eq:KH}
 H_{8\times 8}=\left ( 
 \begin{array}{cc}
 H_c & H_{cv} \\
 H^{\dagger}_{cv} & H_v
 \end{array}
 \right ),
\end{equation}
where $H_c$ is the $2\times 2$ diagonal matrix related to the conduction band ($\Gamma _6$
in the energy band profile - see Fig.~\ref{fig1}(c)] while $H_v$ is the $6\times 6$ diagonal matrix
corresponds to the valence bands ($\Gamma _8$, $\Gamma _7$ in the energy band profile)
\begin{eqnarray}
H_c&=&H_{\Gamma _6}(z)\mathbf{1}_{2\times2},  \\
H_v&=&H_{\Gamma _8}(z)\mathbf{1}_{4\times4} \oplus H_{\Gamma _7}(z)\mathbf{1}_{2\times2}.
\end{eqnarray}
The Hamiltonian $H_{\Gamma _i}(z)$ ($i=6,7,8$) for the band $\Gamma _i$ is expressed as
\begin{equation}
\label{eq:Hi}
H_{\Gamma _i}(z)= -\frac{\hbar ^2}{2m_0} \frac{d^2}{dz^2} + \frac{\hbar ^2 (k_x^2+k_y^2)}{2m_0} +
V_H(z)+V_{\Gamma _i}(z), 
\end{equation}
where $m_0$ is the free electron mass and $V_H(z)$ is the Hartee potential.
The potential energy profile $V_{\Gamma _i}(z)$ in Eq.~(\ref{eq:Hi}) is related to the band-offset
and is
given by
\begin{eqnarray}
 V_{\Gamma _6}(z)&=&h_6(z), \\
 V_{\Gamma _8}(z)&=&-h_8(z)-E_g, \\
 V_{\Gamma _7}(z)&=&-h_7(z)-E_g-\Delta_g, 
\end{eqnarray}
where $h_i(z)=\delta_i h_{QW}(z)+\delta_{bi} h_b(z)$ with $h_{QW(b)}(z)$  being  a dimensionless
functions describing the potential energy profile of the quantum well (central barrier), $\delta
_{i(bi)}$ is the band-offset at the quantum well (central barrier) interface
while $E_g$ and $\Delta_g$ are the energy gap and the split-off band gap, respectively. \\
The off-diagonal element $H_{cv}$ of the Hamiltonian (\ref{eq:KH}) has the form 
\begin{equation}
 H_{cv}=\left ( \begin{array}{cccccc}
         \frac{-\kappa_+}{\sqrt{2}} & \sqrt{\frac{2}{3}}\kappa_z & \frac{\kappa_-}{\sqrt{6}} & 0 &
\frac{-\kappa_z}{\sqrt{3}} & \frac{-\kappa_-}{\sqrt{3}} \\
0 & \frac{-\kappa_+}{\sqrt{6}} & \sqrt{\frac{2}{3}}\kappa_z & \frac{\kappa_-}{\sqrt{2}} &
\frac{-\kappa_+}{\sqrt{3}} & \frac{-\kappa_z}{\sqrt{3}} 
        \end{array}
        \right ),
\end{equation}
where $\kappa _{+,-,z}=Pk_{+,-,z}$, $k_{\pm}=k_x\pm ik_y$ and $P=-i\hbar \langle S|p_x|X \rangle /
m_0$ is the conduction to valence band coupling with $|S\rangle$, $|X \rangle$ being the Bloch
functions at the $\Gamma$ point. \\
Using the folding-down transformation, the $8\times 8$ Hamiltonian (\ref{eq:KH}) can
be reduced into the $2\times 2$ effective Hamiltonian for the
conduction band
\begin{equation}
\label{eq:Hc}
 \mathcal{H}(E)\psi_c=H_c+H_{cv}(E-H_v)^{-1}H_{cv}^{\dagger}.
\end{equation}
Since $E_g$ and $\Delta _g$ are the largest energies in the system, we can expand the on- and
off-diagonal elements of the Hamiltonian (\ref{eq:Hc}) in a series limiting to the first
non-zero elements. This procedure leads to the Hamiltonian 
\begin{eqnarray}
\label{eq:3DH}
 \mathcal{H} &=& \left [ -\frac{\hbar ^2}{2m^*} \frac{d^2}{dz^2} + \frac{\hbar ^2
(k_x^2+k_y^2)}{2m^*} + V_{self}(z) \right ] \mathbf{1}_{2\times 2} \nonumber \\
&+& \alpha(z) 
\left ( 
 \begin{array}{cc}
 0 & k_y+ik_x \\
k_y-ik_x & 0
 \end{array}
 \right ),
\end{eqnarray}
where $V_{self}(z)$ is the self-consistent potential energy profile 
\begin{equation}
\label{eq:vself}
 V_{self}(z)=V_H(z)+\delta _6 h_{QW}(z)+\delta _{b6} h_{b}(z),
\end{equation}
$m^*$ is the effective mass
\begin{equation}
 \frac{1}{m^*}=\frac{1}{m_0}+\frac{2P^2}{3\hbar ^2} \left ( \frac{2}{E_g} + \frac{1}{E_g+\Delta_g}
\right ),
\end{equation}
and $\alpha(z)$ is the Rashba SO coupling constant
\begin{equation}
\label{eq:a}
 \alpha(z)=\alpha_{QW}\frac{dh_{QW}(z)}{dz}+\alpha_{b}\frac{dh_{b}(z)}{dz}-\alpha_{H}\frac{dV_{H
}(z)}{dz}
\end{equation}
with
\begin{eqnarray}
\label{eq:a1}
\alpha _{QW}&=&\frac{P^2}{3} \left [ \frac{\delta _8}{E_g^2} - \frac{\delta _7}{(E_g+\Delta _g)^7}
\right ],  \\
\label{eq:a2}
\alpha _{b}&=&\frac{P^2}{3} \left [ \frac{\delta _{b8}}{E_g^2} - \frac{\delta _{b7}}{(E_g+\Delta
_g)^7}
\right ],  \\
\label{eq:a3}
\alpha _{H}&=&\frac{P^2}{3} \left [ \frac{1}{E_g^2} - \frac{1}{(E_g+\Delta _g)^7}
\right ] .
\end{eqnarray}

\subsection{SO coupling constants}
\label{subsec:SOC}
In this subsection, we briefly describe the procedure used to determine $\alpha(z)$ based on
Eq.~(\ref{eq:a}). The main part of this procedure contains the calculations of the self-consistent 
potential energy profile $V_{self}(z)$ which includes the band potential energy profile, the
potential generated by the gates and doping and the Hartree potential resulting from the
electron-electron interaction. In our calculations, we
start from the single-electron Hamiltonian without the SO interaction and assume that
the electron is confined  in the $z$ direction while in the $x-y$ plane the system is infinite.
This leads to the 1D Schr\"{o}dinger equation in the form
\begin{equation}
\label{eq:RS1D}
\left ( -\frac{\hbar ^2}{2m^*} \frac{d^2}{dz^2} + \frac{\hbar ^2
k_{\parallel}^2}{2m^*} + V_{self}(z) \right )\varphi_n(z)=\mathcal{E} _n\varphi_n(z),
\end{equation}
where $k_{\parallel}^2=k_x^2+k_y^2$.
The eigenproblem (\ref{eq:RS1D}) is solved numerically by the diagonalization in the basis of
infinite quantum well states $\varphi_n(z)=\sum _{j=1}^N c_j \sin(j\pi z/L_z)$, where $L_z$ is the
total length of the heterostructure in the $z$ direction. The Hartee potential $V_H(z)$  [see
Eq.~(\ref{eq:vself})]  is calculated from the Poisson equation
\begin{equation}
\label{eq:poisson}
 \frac{d^2}{dz^2}V_H(z)=-\frac{e}{\epsilon _0 \epsilon _r} [n_e(z)+n_d(z)],
\end{equation}
where $\epsilon$ is the dielectric constant, $n_d(z)$ is the doping profile and $n_e(z)$ is the
electron density, which is given by
\begin{equation}
 n_e(z)=\frac{em^*}{\pi \hbar ^2} k_B T \sum _{n} \ln \left [ 1+e^{(E_F - \mathcal{E} _n
)/k_B T}\right ]
\end{equation}
where $k_B$ is the Boltzmann constant, $T$ is the temperature and $E_F$ is the Fermi energy.
Equation (\ref{eq:poisson}) is solved by the relaxation
method assuming the Dirichlet boundary conditions determined by the gate voltages. In calculations 
we always keep $V_{g1}=0$ as the reference potential.\\  
In the self-consistent procedure, equations (\ref{eq:RS1D}) and (\ref{eq:poisson}) are solved
iteratively until the convergence is reached. The self-consistent potential energy profile and the
corresponding wave functions for two lowest states $\varphi_1$ and $\varphi _2$ are presented in
Fig.~\ref{fig1}(d). Then, the SO coupling $\alpha (z)$ is determined from the potential
$V_{self}(z)$   by the use of Eq.~(\ref{eq:a}).
The present calculations have been performed for 
Al$_{0.48}$In$_{0.52}$As/Ga$_{0.47}$In$_{0.53}$As double quantum well with the
following material parameters:\cite{Vurgaftman2001} $E_g=0.8161$~eV, $\Delta
_g=0.3296$~eV, $\delta _6=0.52$~eV, $\delta _{7}=0.1637$~eV, $\delta _{8}=0.1935$~eV,$\delta
_{b6}=0.21$~eV, $\delta _{b7}=0.1343$~eV, $\delta _{b8}=0.152$~eV, $m^*=0.043$ and $E_P=2m_0P^2 /
\hbar ^2 =25.3$~eV. The dielectric constant $\epsilon _r=14.013$ is assumed to be
constant in the entire heterostructure.

\subsection{Effective 2D Hamiltonian and conductance calculations}
Now, we derive an effective 2D Hamiltonian in the two subband model starting from its 3D version
given by Eq.~(\ref{eq:3DH}). For this purpose we define the four element basis $\{
|\varphi _1,\uparrow \rangle |\varphi _1,\downarrow \rangle, |\varphi _2,\uparrow \rangle, |\varphi
_2,\downarrow \rangle\}$ which consists of
the spin-degenerate ground and first excited eigenstate of the Hamiltonian (\ref{eq:RS1D})
without SO interaction. The projection of (\ref{eq:3DH}) onto this basis leads to the $4 \times 4$
Hamiltonian given by
\begin{widetext}
\begin{equation}
\label{eq:2DH}
 \mathcal {H} _{2D}= \left ( \begin{array}{cccc}
 \frac{\hbar ^2 k _{\parallel}^2}{2m^*} + \varepsilon _1 & \alpha _{11} (k_y+ik_x) & 0 &
\alpha_{12}(k_y+ik_x) \\
 \alpha _{11} (k_y-ik_x) & \frac{\hbar k _{\parallel}^2}{2m^*} + \varepsilon_1 & 
\alpha_{12}(k_y-ik_x) & 0 \\
 0 & \alpha_{12}(k_y+ik_x)  & \frac{\hbar ^2 k _{\parallel}^2}{2m^*} + \varepsilon_2 & \alpha
_{22} (k_y+ik_x)\\
 \alpha_{12}(k_y-ik_x) & 0 & \alpha _{22} (k_y-ik_x) & \frac{ \hbar ^2 k
_{\parallel}^2}{2m^*} +
\varepsilon_2  
 \end{array}
 \right )
\end{equation}
\end{widetext}
where $\alpha _{nm} = \langle \varphi _n| \alpha (z) | \varphi _m \rangle$ with $n,m=1,2$.  

The calculations of the  conductance have been performed within the scattering matrix
method using the Kwant package.~\cite{kwant} For this purpose we have transformed the
Hamiltonian (\ref{eq:2DH}) into the discretized form on the grid $(x_{\mu}, y_{\nu})= \mu dx, \nu 
dx$ ($\mu, \nu = 1,2, \ldots$) where $dx$ is the lattice constant.
We introduce the discrete representation of the electron state in the $4 \times 4$ space as follows:
$|\Psi(x_{\mu}, y_{\nu})\rangle
=
\left(|\psi_1^{\uparrow}( x_{\mu},y_{\nu})\rangle
,|\psi_1^{\downarrow}( x_{\mu},y_{\nu})\rangle, |\psi_2^{\uparrow}( x_{\mu},y_{\nu})\rangle
|\psi_2^{\downarrow}( x_{\mu},y_{\nu})\rangle \right)^T
= |\Psi_{\mu, \nu}\rangle$.
Introducing a set $\boldsymbol{\tau}$ of Pauli-like matrices in the orbital space, the Hamiltonian
(\ref{eq:2DH}) takes on the discretized form 
\begin{eqnarray}
\mathcal{H}_{2D}&=& \sum\limits_{\mu\nu}  \left [ (4t + \varepsilon _+) \mathbf{1}
\otimes \mathbf{1} - \varepsilon _- \tau _z \otimes \mathbf{1}  \right ] |
\Psi_{\mu,\nu } \rangle \langle \Psi_{\mu,\nu}| \nonumber \\
 &+& \sum _{{\mu}{\nu}} \bigg \{ t \mathbf{1} \otimes \mathbf{1} + it_{SO} \bigg [ \alpha_{11}
\frac{1}{2}(\mathbf{1}-\tau _z)\otimes \sigma_y \nonumber \\
&+& \alpha_{22} \frac{1}{2}(\mathbf{1}+\tau _z)\otimes \sigma_y + \alpha _{12}\tau _x \otimes
\sigma _y \bigg ] \bigg \} + H.c.\nonumber \\
 &+& \sum _{{\mu}{\nu}} \bigg \{ t \mathbf{1} \otimes \mathbf{1} + it_{SO} \bigg [ \alpha_{11}
\frac{1}{2}(\mathbf{1}-\tau _z)\otimes \sigma_x \nonumber \\
&+& \alpha_{22} \frac{1}{2}(\mathbf{1}+\tau _z)\otimes \sigma_x + \alpha _{12}\tau _x \otimes
\sigma _x \bigg ] \bigg \} + H.c.
 \label{HTB}
\end{eqnarray}
where $t=\hbar ^2/(2m dx^2)$, $t_{SO}=1/(2dx)$ and $\mathbf{1}$ is the $2 \times 2$ unity matrix. 

Let us assume that the electron with spin up in the first subband  is injected from the source
(polarizer) into the conduction channel. The electron can be transmitted via the conduction channel
to the analyzer in one of the four possible processes: (i) intra-subband transmission with spin
conservation $(T_{11}^{\uparrow \uparrow})$, (ii) intra-subband transmission with spin-flip
$(T_{11}^{\uparrow \downarrow})$, (iii) inter-subband transmission with spin conservation
($T_{12}^{\uparrow \uparrow}$) and (iv) inter-subband transmission with spin flip ($T_{12}^{\uparrow
\downarrow}$), where $T_{nm}^{\sigma \sigma'}$ with $\sigma, \sigma'=\uparrow, \downarrow$ and
$n,m=1,2$ denotes the probabilities of the transmission processes (i) - (iv). Similar
scattering processes can be introduced for the spin-up electrons injected from the second
subband. Their probabilities are denoted by $T_{22}^{\uparrow \uparrow}$, $T_{22}^{\uparrow
\downarrow}$, $T_{21}^{\uparrow \uparrow}$, $T_{21}^{\uparrow \downarrow}$. \\

Having determined the transmission coefficients $T^{\sigma \sigma '}_{nm}$
we calculate the conductance in the ballistic regime using the Landauer formula
\begin{equation}
 G_{nm}^{\sigma \sigma^{\prime}}=\frac{e^2}{h} \int T_{nm} ^{\sigma \sigma ^{\prime}} (E) \left (
\frac{\partial f_{FD}(E,E_F)}{\partial E} \right ) dE,
\end{equation}
where $\sigma$, $\sigma^{\prime}$ are the spin indices and $f_{FD}(E,E_F)=1/[1+\exp(E-E_F)/k_BT]$ is
the Fermi-Dirac distribution function, where $T$ is the temperature and $E_F$ is the Fermi energy.

For the assumed 100\% spin injection (detection)
efficiency of the contacts, the total conductance via the device is
given by 
\begin{equation}
 G=\sum _{n,m=1}^{2} G_{nm}^{\uparrow \uparrow}.
\end{equation}
The conductance calculations presented in the paper have been performed for $dx=2$~nm and
$T=4.2$~K.

\section{Results}
\label{sec3}
In this section we study the conductance through the spin transistor including 
the intra- and inter-subband SO interaction. We start from
the model, in which the SO coupling constants are treated as the parameters (subsection~A)
and show how the inter-subband-induced SO coupling affects the spin transistor operation. Then, in
subsection B, we introduce the realistic model with the
Al$_{0.48}$In$_{0.52}$As/Ga$_{0.47}$In$_{0.53}$As double quantum well, for which the SO coupling
constants are determined by the Schr\"{o}dinger-Poisson approach
presented in subsec.~\ref{subsec:SOC}. 
\subsection{Parametrized model}
We consider the spin transistor with the length $L=800$~nm and the gate attached to the conduction
channel in the middle of the nanostructure. The length of the gate $L_g=400$~nm  (see
Fig.~\ref{fig1}).
The energy difference between the two subbands is taken to be $\Delta \varepsilon = \varepsilon _2 -
\varepsilon_1=1$~meV [Eq.~(\ref{eq:2DH})]. We assume the channel width $W=40$~nm which guaranteers
that the energy separation between the two lowest energy states related to the confinement in the
lateral $y$ direction $\Delta \varepsilon _{\perp} \approx \hbar ^2 \pi ^2 / 2 m^* W^2=4.2$~meV is 
greater than $\Delta \varepsilon$. All results presented in this subsection have been obtained for
the Fermi energy $E_F=4$~meV, which ensures that only the lowest energy state in the transverse
motion ($y$ direction) is occupied and the two subbands in the grown $z$ direction participate in
the transport. The SO coupling constants, experimentally controlled by the gate voltage, are
treated as the parameters of the calculations. \\

\begin{figure}[ht]
\begin{center}
\includegraphics[scale=0.40]{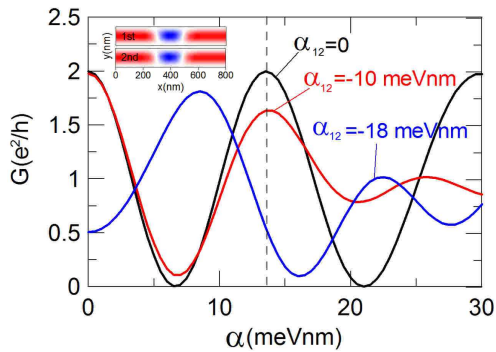}
\caption{Conductance $G$ as a function of intra-subband SO coupling constant $\alpha$ for
different inter-subband couplings $\alpha _{12}$. Vertical dashed line marks the value of
$\alpha$, for which the spin density distributions are depicted in Fig.~\ref{fig5}. Inset: spin
density distributions $s^{1(2)}_z$ for the first (1st) and second (2nd) subband for
$\alpha=14$~meVnm and $\alpha _{12}=0$.}
\label{fig2}
\end{center}
\end{figure}

Let us start our study from the case in which the intra-subband SO coupling constants, in the both
subbands are equal $\alpha _{11}=\alpha _{22}= \alpha$. Figure~\ref{fig2} presents the conductance
as a function of $\alpha$ for different inter-subband SO coupling constant $\alpha
_{12}$. We assume that $\alpha _{12}$ takes on the negative values which is consistent with the
results for the realistic structure (see subsection~\ref{sec3:real}). As we have checked,
the change of sign $\alpha _{12}$ does not change the conductance in any way -- the conductacne
depends only on the absolute value of $\alpha _{12}$.
For the inter-subband SO coupling constant $\alpha _{12}=0$ the spin dynamics in the two
subbands, via which the electrons are transmitted, is independent. The spin of electron flowing
in the vicinity of the gate rotates due to the SO interaction. Since the intra-subband SO coupling
are assumed to be equal, the electron spin in each of the subbands precesses with
the same precession length $\lambda _{SO}=2\pi/\Delta k$, where
$\Delta k=k_F^{\uparrow}-k_F^{\downarrow}=2m^*\alpha /\hbar ^2$. In this case, the slight
difference in the transport conditions through the both subbands can result from the energy
difference $\Delta \varepsilon$, however it is too small to affect the spin transistor operation.
Hence, if we assume the ideal spin polarizer (analyzer), which transmits only electrons with well
defined spin, the conductance oscillates as a function of $\alpha$ according to the
formula~\cite{Spintronics}
\begin{equation}
\label{eq:Ga0}
G=2G_0 \cos ^2 \left ( \frac{\Delta k L_g}{2} \right )=2G_0 \cos ^2 \left ( \frac{m^* \alpha
L_g}{\hbar ^2} \right ),
\end{equation}
where $G_0=e^2/h$ and the factor $2$ is related to the fact that the electrons are
transmitted via the two subbands.\\
These conductance oscillations as a function of the intra-subband SO coupling constant $\alpha$ are
presented in Fig.~\ref{fig2} (black line, $\alpha _{12}=0$). Based on Eq.~(\ref{eq:Ga0}) one can
conclude that the conductance reaches maximum for $\Delta k L_g  =  2 N \pi$, which
corresponds to the process, in which the spin of the electron flowing through the conduction
channel precesses the integer number of times. On the other hand, the conductance minimum is reached
for  $\Delta k L_g = (2 N+1) \pi$, which corresponds to the half-integer rotation number of the
electron spin. The former case is depicted in the inset of Fig.~\ref{fig2}, in which we present the
spin density distributions
$s^{1(2)}_z$ in the nanostructure calculated for the both subbands  for $\alpha=14$~meV and $\alpha
_{12}=0$.
\begin{figure}[ht]
\begin{center}
\includegraphics[scale=0.45]{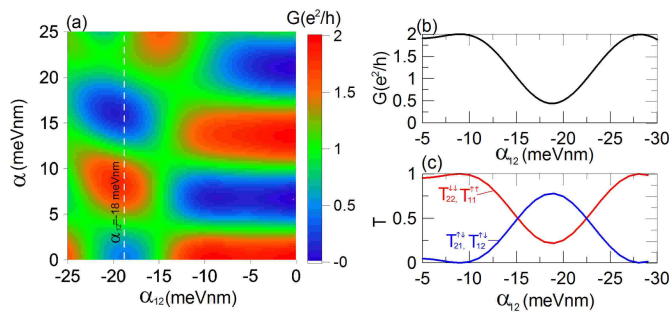}
\caption{(a) Conductance $G$ as a function of intra-subband SO coupling constant $\alpha$ and
inter-subband spin-orbit coupling constant $\alpha _{12}.$ (b) Conductance $G$ and (c)
transmission probabilities as a function of $\alpha_{12}$ for $\alpha=0$.}
\label{fig3}
\end{center}
\end{figure}

The regular oscillations of $G(\alpha)$ are modified if we introduce the inter-subband SO coupling
into the system, i.e. $\alpha _{12}\ne0$. As presented in Fig.~\ref{fig2}, for $\alpha
_{12}=-10$~meVnm the change of the conductance becomes significant for large values of $\alpha$. The
black and  red line come together for small  $\alpha$ and diverge for $\alpha>10$~meVnm.
For $\alpha=14$~meVnm, marked by the vertical dashed line, the inter-subband SO
interaction leads to the slight reduction of the conductance. The further change of the
inter-subbands SO coupling constant up to $\alpha _{12}=-18$~meVnm leads to the inversion of
the oscillations, namely the conductance reaches the minimum for $\alpha$ for which it is maximal in
the case of $\alpha _{12}=0$. This inversion is clearly visible in Fig.~\ref{fig3} which presents
the conductance as a function of the intra- and inter-subband SO coupling constants $G(\alpha,
\alpha _{12})$. The complete inversion is observed for $\alpha _{12}=-18$~meVnm (white dashed line),
for which also the period of the $G(\alpha)$ oscillations
slightly increases. Interestingly, as presented in Fig.~\ref{fig3}(b) even for 
$\alpha=0$ corresponding to the symmetric heterostructure, the conductance oscillates as a
function of $\alpha _{12}$. The transmission probabilities
shown in Fig.~\ref{fig3}(c) indicate that this behavior is directly related to the increase of the
inter-subbands spin-flip transmission probability. All these results suggest the possible
application of the inter-subband SO interaction in the spin transistor design, although the
experimental control of $\alpha _{12}$ still remains an open issue.

\begin{figure}[ht]
\begin{center}
\includegraphics[scale=0.35]{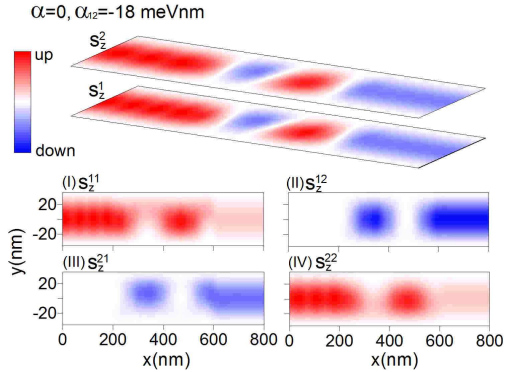} \\
\caption{Spin density distribution $s^{1(2)}_z$ for the 1st and 2nd subband (upper panels)
calculated for $\alpha=0$ and $\alpha _{12}=-18$~meVnm. Figures (I) and (II) correspond to
the spin density distribution in the 1st and 2nd subband, respectively, if the electron with spin
up is injected into the first subband, while figures (III) and (IV) correspond to the spin
density distribution in the 1st and 2nd subband if the electron with spin up is injected into the
2nd subband.}
\label{fig4}
\end{center}
\end{figure}

The conductance behavior [Figs.~\ref{fig2} and \ref{fig3}] result from the spin dynamics, which in
the presence of the inter-subband SO interaction becomes much more complicated. Similarly, as for
$\alpha _{12}=0$, the spin dynamics is determined by the differences of $k_F$ Fermi wave vector
between the subbands participating in the transport for the given Fermi energy. These
differences can be determined from the eigenenergies of Hamiltonian (\ref{eq:2DH}), which are given
by
\begin{equation}
E_{ks\rho}= E_0+\frac{h^2 k^2}{2m^*}+\varepsilon _+ + s \alpha _+ k +\rho \sqrt{(\alpha
_{12}k)^2+(\varepsilon _- + s \alpha _- k)^2},
\end{equation}
where
\begin{equation}
 \varepsilon _{\pm}=\frac{1}{2} (\varepsilon _1 \pm \varepsilon_2), \:\:\:\:
 \alpha _{\pm}=\frac{1}{2} (\alpha _1 \pm \alpha _2),
\end{equation}
$E_0$ is the energy of the lowest state related to the confinement in the lateral $y$ direction
while $s=\pm 1$ and $\rho=\pm1$ correspond to the spin state and the subband, respectively. \\
Notice, that the electron
initially injected into the channel within spin up state oscillates between the subbands changing
its spin. The spin dynamics is the combination of the precession with different precession
lengths which, in contrast to the case with $\alpha _{12}=0$, depend on the Fermi energy. The
simplest case for which this problem can be solved
analytically is the symmetric structure with zero intra-subband SO coupling ($\alpha=0$),
presented in Fig.~\ref{fig3}(b). Then, the spin precession length is given by 
\begin{equation}
 \lambda _{SO}=\frac{2 \pi}{\Delta k}=\frac{2 \pi}{k_2 - k_1}, 
 \label{eq:precess}
\end{equation}
where
\begin{widetext}
\begin{eqnarray}
\label{eq:k1}
k_1=\frac{\sqrt{2m^*E_F}}{\hbar} \sqrt{1-\frac{\varepsilon _{+}}{E_F}+\frac{\alpha _{12}^2m^*}{\hbar
^2 E_F} \left ( 1+\sqrt{1+\frac{2\hbar ^2 (E_F-\varepsilon _{-})}{\alpha _{12}^2m^*} +\frac{\hbar
^4 \varepsilon _{-}^2}{\alpha _{12}^4m^{*2}}}\right ) }, \\
\label{eq:k2}
k_2=\frac{\sqrt{2m^*E_F}}{\hbar} \sqrt{1-\frac{\varepsilon _{+}}{E_F}+\frac{\alpha _{12}^2m^*}{\hbar
^2 E_F} \left ( 1-\sqrt{1+\frac{2\hbar ^2 (E_F-\varepsilon _{-})}{\alpha _{12}^2m^*} +\frac{\hbar
^4 \varepsilon _{-}^2}{\alpha _{12}^4m^{*2}}}\right ) }. 
\end{eqnarray}
\end{widetext}
In Fig.~\ref{fig4} we present the $z$ component  $s^{1(2)}_z$ of the spin density distribution for
the both subbands and $\alpha _{12}=-18$~meVnm corresponding to the conductance minimum in
Fig.~\ref{fig3}(b). The lower panels in Fig.~\ref{fig3} depict 
the partial spin density distributions: (I) $s_z^{11}$ and (II) $s_z^{12}$ correspond to
the spin density distribution in the first and second subband, respectively, if the electron with
spin up is injected into the first subband, while $s_z^{21}$ (III) and $s_z^{22}$ (IV) correspond
to the spin density distribution in the first and second subband, if the electron with spin up is
injected into the second subband. These partial spin density distributions give us information
not only about the spin dynamics in the considered subband but also about the spin behavior due to
the inter-subband transitions.
\begin{figure}[ht]
\begin{center}
\includegraphics[scale=0.35]{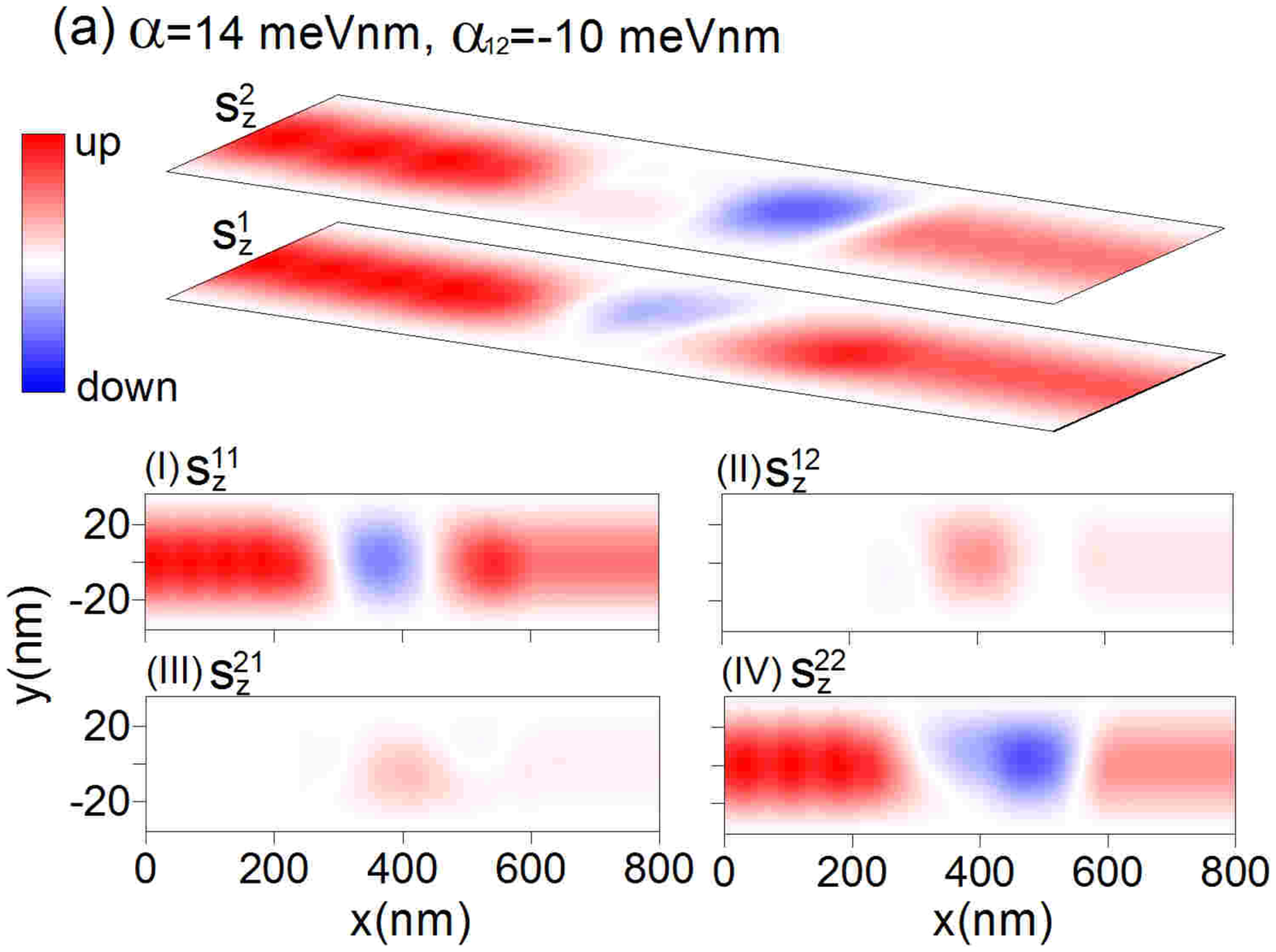} \\
\vspace{5mm}
\includegraphics[scale=0.35]{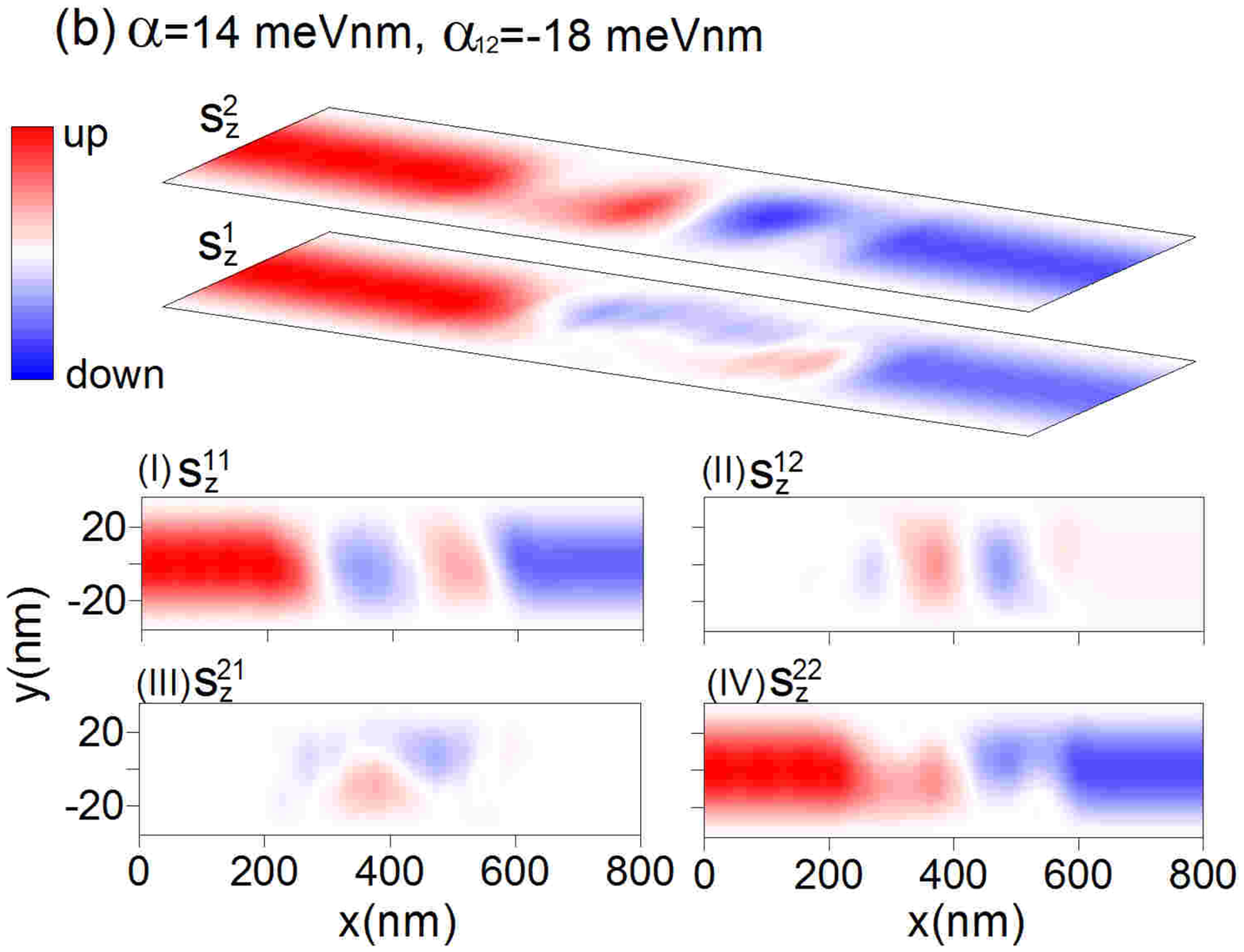} 
\caption{Spin density distribution $s^{1(2)}_z$ for the 1st and 2nd subband (upper panels)
calculated for $\alpha=14$~meVnm and (a) $\alpha _{12}=-10$~meVnm, (b)
$\alpha _{12}=-18$~meVnm. Figures (I) and (II) correspond to the spin density distribution in
the 1st and 2nd subband, respctively if the electron with spin up is injected into the first
subband, while figures (III) and (IV) correspond to the spin density distribution in
the 1st and 2nd subband if the electron with spin up is injected into the 2nd subband.}
\label{fig5}
\end{center}
\end{figure}
For the both subbands (Fig.\ref{fig4}) the electrons initially injected with spin up
reverse their spin when flowing through the nanodevice [cf. $s^1_z$ and $s^2_z$].  The spin-down
electrons reaching the output are backscattered from the ideal spin-up polarized contact (analyzer),
which leads to the decrease of the conductance presented in Fig.~\ref{fig3}(b). Since the
intra-subband SO coupling constant $\alpha=0$ the spin of the electron flowing through the subband,
in which it was injected,  does not precess [cf. Figs.~\ref{fig3} (I) and (IV)].  Nevertheless, as
presented in Fig.~\ref{fig3} (II) and (III) the electron spin is inverted in the inter-subband
transitions, the probability of which reaches maximum for the chosen $\alpha _{12}$. In this
case, the spin precession length is given by Eq.~(\ref{eq:precess}).

The spin dynamics near the gate becomes more complicate for the nonzero intra-subband SO coupling
($\alpha \ne 0$). In this case the spin degeneracy of the subbands is lifted. Thus, beside the
reversal of spin related to the inter-subband transition, we expect the intra-subband spin
precession. In Fig.~\ref{fig5} we present the spin density
distributions for $\alpha=14$~meVnm (marked by the vertical dashed line in Fig.~\ref{fig2}) and two
chosen values of the inter-suband SO coupling constants (a) $\alpha _{12}=-10$~meVnm and (b) $\alpha
_{12}=-18$~meVnm. As shown in Fig.~\ref{fig5}, for $\alpha _{12}=-10$~meVnm the spin
dynamics in the nanostructure differs only slightly from the case without the inter-subband SO
interaction (compare with the inset of Fig.~\ref{fig2}). The electron spin performances one full
rotation and leaves the nanodevice with almost the same spin as on the input.  In contrast to the
case with $\alpha _{12}=0$, we observe the inter-subband transition in which the electron
conserves its spin. The spin dynamics drastically changes if we increase the inter-subband SO
coupling. For $\alpha _{12}=-18$~meVnm [Fig.~\ref{fig5}(b)] the electrons initially injected with
spin up reverse their spin on the output leading to the decrease of the conductance. Note that
in contrast to the spin dynamics for $\alpha=0$, for which the spin
flip is related to the inter-subband transition, in the case of $\alpha \ne 0$ the electron
conserves the spin in this type of transitions. Due to the intra-subband SO interaction the spin
precession takes place mainly
in the subband into which the electron is injected. However this precession is strongly
affected by the inter-subband SO interaction, which significantly changes the precession length.
\begin{figure}[ht]
\begin{center}
\vspace{5mm}
\includegraphics[scale=0.4]{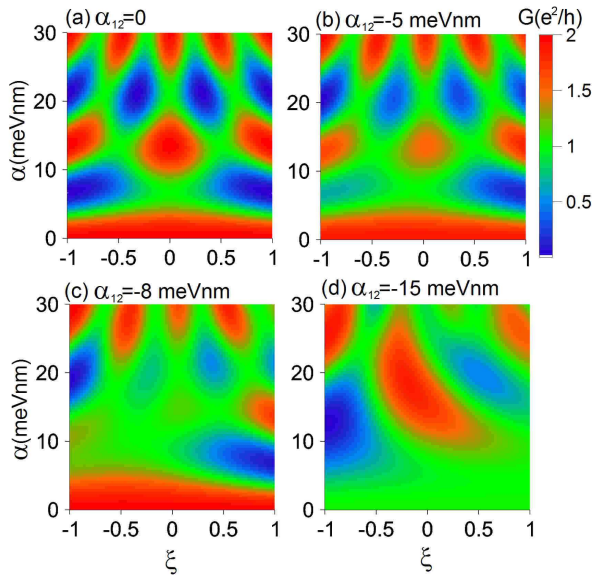}
\caption{Conductance $G$ as a function of intra-subband SO coupling constant $\alpha$ and 
asymmetry parameter $\xi=\alpha_{11} / \alpha_{22}$ for different  $\alpha
_{12}$.}
\label{fig6}
\end{center}
\end{figure}

Finally, we have also performed the calculations of the conductance for the most general
case for which the intra-subband spin-orbit coupling constants are different in the both
subbands. For this purpose
we define the parameter of the asymmetry $\xi=\alpha_{11}/\alpha _{22}$. Fig.~\ref{fig6}
displays the conductance as a function of the intra-subband SO coupling constant $\alpha$ and the
asymmetry parameter $\xi$ for different $\alpha _{12}$. We 
see that even in the absence of the inter-subband SO interaction [Fig.~\ref{fig6}(a)] the asymmetry
of the intra-subband SO coupling strongly affects the conductance oscillations making them
irregular. In this case the conductance is symmetric relative to the subband interchange. As shown
in Figs.~\ref{fig6}(b)-(d) this symmetry is lifted by the inter-subband SO interaction.

\subsection{Realistic model}
\label{sec3:real}
\begin{figure*}[ht]
\begin{center}
\includegraphics[scale=0.7]{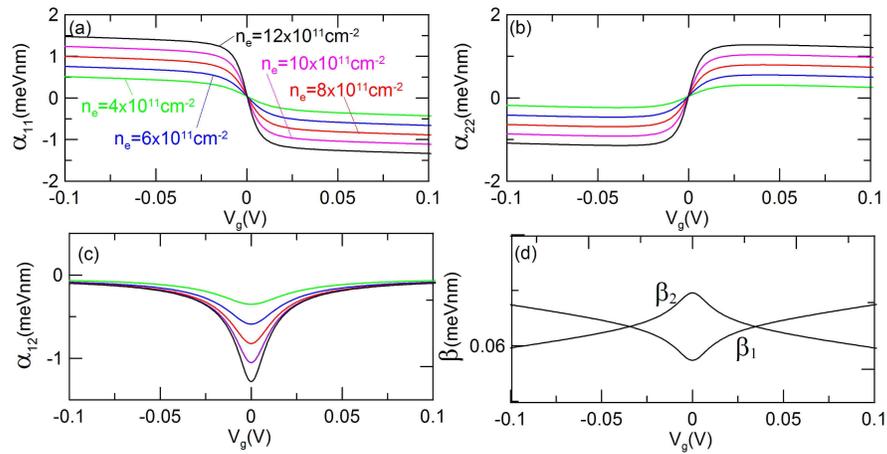}
\caption{Intra-subband (a) $\alpha _{11}$ and (b) $\alpha _{22}$ and inter-subband (c) $\alpha
_{12}$ SO coupling constants as a function of gate voltage $V_g$ for different electron
densities $n_e$.(d) Dresselhaus SO coupling constants  as a function of  gate voltage $V_g$ for
$n_e=10 \times 10
^{11}$~cm$^{-2}$.}
\label{fig7}
\end{center}
\end{figure*}

In this subsection we study the conductance of the spin transistor with the
conduction channel formed from the Al$_{0.48}$In$_{0.52}$As/Ga$_{0.47}$In$_{0.53}$As double quantum
well presented in Fig.~\ref{fig1}(b). For this heterostructure we have determined the SO
coupling constants using the self-consistent procedure described in subsec.~\ref{subsec:SOC}.
Figure~\ref{fig7} presents the intra- ($\alpha _{11}$ and $\alpha _{22}$) and inter-subband ($\alpha
_{12}$) Rashba SO coupling constants as a function of the gate voltage for different electron
concentrations. We have also calculated the Dresselhaus SO coupling defined as
\begin{equation}
 \beta _n=\beta ^{3D} \langle \varphi _n| \hat{k}_z^2 | \varphi _n \rangle,
\end{equation}
where $\beta ^{3D}$ is the Dresselhaus SO coupling for the bulk taken on as $\beta
^{3D}=0.0237$~meVnm$^3$.\cite{Jancu2005} Figure \ref{fig7} shows that for the considered wide
quantum well, the Dresselhaus SO coupling constants $\beta _{1(2)}$ are two orders of magnitude
smaller than the Rashba constants [Fig.~\ref{fig7}(d)]. Therefore, the Dresselhaus SO interaction
is neglected in the conductance calculations presented in the rest of the paper.

Figure~\ref{fig7}(c) shows that the inter-subband SO coupling constant is an even function of the
gate voltage and exhibits the "resonant behavior" around $V_g=0$ corresponding to the symmetric
geometry of the heterostructure.
Simultaneously, at the resonant voltage $V_g=0$, the intra-subband SO coupling constants $\alpha
_{11}$ and $\alpha _{22}$ change the sign.
Similar "resonant behavior" was recently reported by Calsaverini et. al. for
InSb/Al$_{0.12}$In$_{88}$Sb double quantum well.~\cite{Calsaverini2008} The
authors~\cite{Calsaverini2008} argued that this feature results from the dominant role of
the Hartree potential and the overlap between the wavefunctions of the ground and the first excited
state in the quantum well, which for $V_g=0$ becomes maximal. Notice
that the conduction channel, in which the SO coupling constants rapidly change around
$V_g=0$ is preferred for the application in the spin transistor architecture in which the switching
between the on and off states should be realized in the gate voltage range as narrow as possible
[see Fig.~\ref{fig1}(e)]. We have performed the calculations of Rashba constants for different
electron densities (Fig.~\ref{fig7}) taking care that only the two lowest-energy states in the
quantum well were occupied. As shown in Fig.~\ref{fig7} the increasing
electron density $n_e$ leads to the increase of the slope $\alpha_{11(22)}(V_g)$ curves around
$V_g=0$ making the heterostructure more convenient for the spin transistor application.
Simultaneously, the inter-suband SO coupling $\alpha _{12}$ at $V_g=0$ decreases which, as we will
show later, also affects the conductance at this gate voltage. 

\begin{figure}[ht]
\begin{center}
\includegraphics[scale=0.4]{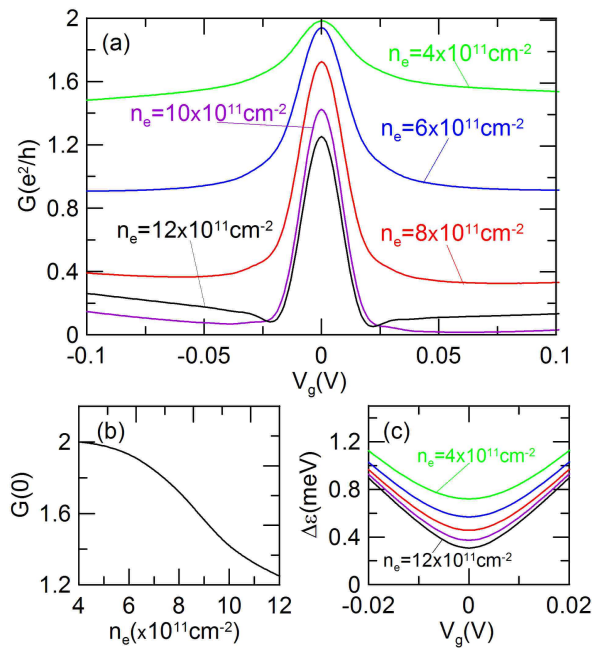}
\caption{(a) Conductance $G$ as a function of gate voltage $V_g$ for different electron
densities $n_e$. (b) Conductance $G(0)$ for $V_g=0$ as a function of electron density
$n_e$. (c) Energy difference between the subbands $\Delta \varepsilon = \varepsilon_2 -
\varepsilon_1$ as a function of gate voltage $V_g$ for the same electron densities as in
figure (a).}
\label{fig8}
\end{center}
\end{figure}

Having the SO coupling constants determined from the Schr\"{o}dinger-Poisson approach we calculate
the conductance using of the scattering matrix method. For this purpose we consider the spin
transistor with length $L=3$~$\mu$m, width $W=40$~nm and the gate located in the middle of
the conduction channel. The length of the gate $L_g=2$~$\mu$m is assumed to be comparable to that
used in recent experiment.~\cite{Chuang2015} Figure~\ref{fig8}(a) depicts the conductance as a
function of the gate voltage for different electron densities. 
The conductance $G(V_g)$ exhibits the pronounced peak around $V_g=0$
related to the low resistance state of the spin transistor (on state). The change of the gate
voltage in the narrow range around $V_g=0$ switches the transistor into the low conductance state
with the high resistance (off state). Notice that the on/off conductance ratio strongly depends
on the electron density and is larger for high $n_e$. The dependence $G(V_g)$ is
determined by the spin dynamics in the conduction channel, which depends on the strength of the
Rashba SO interaction. At $V_g=0$ related to the symmetric geometry of the heterostructure, the
intra-subband SO coupling constants
$\alpha _{11}=\alpha _{22}=0$ [see Fig.~\ref{fig7}(a)(b)]. Then, in the absence of the inter-subband
SO interaction, the spin of the electron injected from the polarizer does not precess and the
electron leaves the conduction channel with the same spin matching the polarization of the left
contact (analyzer). The both subbands
transmit the electrons giving raise to the
conductance $G=2 e^2 /h$. However, as shown in Fig.~\ref{fig7}(a) and (b) the slight deviation of
the gate voltage from $V_g=0$ causes the rapid change of the intra-subband SO coupling constants.
In particular, if the strength of this SO interaction is sufficient to inverse the spin of the
electron flowing through the nanostructure the electron is reflected from the analyzer,
which results in the zero conductance. As shown in Fig.~\ref{fig8}(a) the large changes of the
conductance around $V_g=0$ are strictly related to the abrupt change of the SO coupling constants
presented in Fig.~\ref{fig7}. Outside the close vicinity of $V_g=0$  the conductance is almost
constant, which results from the nearly constant values of the SO coupling in this range
(see Fig.~\ref{fig7}). 
\begin{figure}[ht]
\begin{center}
\includegraphics[scale=0.4]{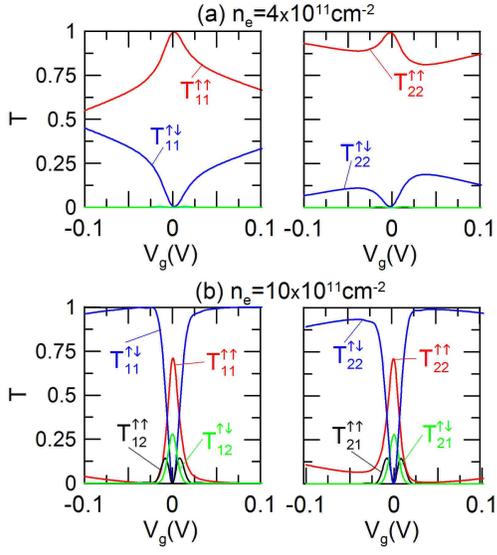}
\caption{Transmission probabilities $T$ as a function of gate voltage $V_g$ for the electron
density (a) $n_e=4\times 10^{11}$~cm$^{-2}$ and (b) $n_e=10\times 10^{11}$~cm$^{-2}$. }
\label{fig9}
\end{center}
\end{figure}

The model of spin dynamics presented above is correct in the absence of the
inter-subband-induced SO interaction or, for the realistic structure, out of the range of the
conductance peak where the inter-subband SO coupling constant is much smaller than the intra-subband
coupling constants. However, in the gate voltage range, in which the conductance peak occurs,
i.~e. around $V_g=0$, the inter-subband SO interaction pays a significant role. The strong evidence
of this interaction is the value of the conductance for $V_g=0$. 
\begin{figure}[ht]
\begin{center}
\includegraphics[scale=0.4]{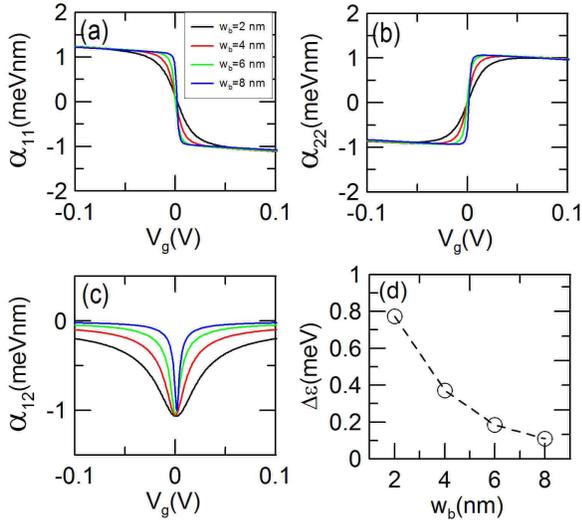}
\caption{Intra-subband (a) $\alpha _{11}$ and (b) $\alpha _{22}$ and inter-subband (c) $\alpha
_{12}$ SO coupling constants as a function of gate voltage $V_g$ for different barrier widths
$w_b$. (d) Energy separation between the subbands $\Delta \varepsilon = \varepsilon_2 -
\varepsilon_1$ at $V_g=0$ as a function of barrier width $w_b$. Results for $n_e=10 \times 10
^{11}$~cm$^{-2}$.}
\label{fig10}
\end{center}
\end{figure}
As mentioned before for $V_g=0$ for which $\alpha _{11}=\alpha _{22}=0$, the absence of the
inter-subband SO interaction leads to $G(0)=2e^2 /h$. However, as depicted in Fig.~\ref{fig8}(a)
this value of the conductance is reached only for the low electron density $n_e=4 \times 10
^{11}$~cm$^{-2}$ for which the inter-subband SO coupling is low (see Fig.~\ref{fig7}). For higher
electron densities, $G(0)$ gradually decreases leading to the reduction of the on/off conductance
ratio. As shown in Fig.~\ref{fig4}, for $V_g=0$ the only possible
process, which decreases the conductance, is the inter-subband transmission with spin-flip resulting
from the inter-subband SO interaction. The probability of this process depends not only on the value
of $\alpha _{12}$ but also on the energy separation between the subbands $\Delta \varepsilon
=\varepsilon_2 - \varepsilon_1$. In Fig.~\ref{fig8}(c), we present $\Delta \varepsilon$ versus
$V_g$. Comparing the results of Fig.~\ref{fig7}(c) and Fig.~\ref{fig8}(c), we see that
for increasing $n_e$, $\alpha _{12}$ increases while  $\Delta \varepsilon$ decreases. These 
effects all together enhance the inter-subband transition with spin flip and leads to the
conductance reduction at $V_g=0$. In order to show this, in Fig.~\ref{fig9} we present the
transmission probabilities as a function of the gate voltage for electron densities (a)
$n_e=4\times 10 ^{11}$~cm$^{-2}$ and (b) $n_e=10\times10^{11}$~cm$^{-2}$.
\begin{figure}[ht]
\begin{center}
\includegraphics[scale=0.4]{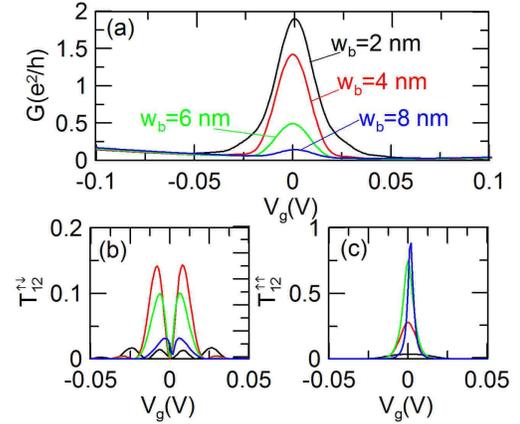}
\caption{(a) Conductance $G$ as a function of gate voltage $V_g$ for different barrier widths
$w_b$. (b) Inter-subband transmission with spin conservation $T_{12}^{\uparrow \uparrow}$ and (c)
inter-subband transmission with spin flip $T_{12}^{\uparrow \downarrow}$ as a function of gate
voltage $V_g$.}
\label{fig11}
\end{center}
\end{figure}
For the low electron density $n_e=4\times10^{11}$~cm$^{-2}$  the
inter-subband transmissions is absent both for electrons injected from the first and the second
subband. The decrease of the conductance for $V_g \ne 0$ corresponds to the increase of the
intra-subband transmission with spin flip. For the high electron density
$n_e=10\times 10^{11}$~cm$^{-2}$ the probability of the inter-suband transmissions is nonzero
around $V_g=0$. Notice that at $V_g=0$ the inter-subband transmission always accompanies the
spin flip while the probability of the inter-subband transmission with spin conservation
$T_{12}^{\uparrow \uparrow}=T_{21}^{\uparrow \uparrow}=0$. It is worth mentioning  that the
transmission probabilities for positive and negative gate voltages are not equivalent leading to the
nonsymmetric dependence $G(V_g)$ presented in Fig.~\ref{fig8}. This asymmetry emerges for high
gate voltages for which the inter-subband SO interaction is weak. Hence, it results from
the asymmetry of the intra-subband SO coupling constants for the ground and first excited state
which is analogous to that observed in Fig.~\ref{fig6}.

As shown above the conductance in the vicinity of $V_g=0$ is mainly determined by the inter-subband
transitions which emerge in the system as a result of the inter-subband SO interaction. This leads
to the question how the width of the central barrier $w_b$, which directly determines the coupling
between the quantum wells, affects the conductance in the considered gate voltage range. In
Fig.~\ref{fig10}~(a)-(c) we present the intra- and inter-subband SO
couplings as a function of the gate voltage calculated for different barrier widths.
Figure~\ref{fig10}(a) shows  that the resonant behavior of $\alpha _{12}$ is more pronounced
for the wide central barrier, while the width of the barrier almost does not change the value of
$\alpha _{12}$ at $V_g=0$. 
In addition, the slopes $d \alpha _{11} / dV_g$ and $d \alpha _{22} / dV_g$ at $V_g=0$ increase
with the increasing $w_b$ making the system more suitable for the spin transistor application.
However, as presented in Fig.~\ref{fig11}(a), the conductance at $V_g=0$ is strongly reduced for
the wide barrier giving raise to the low on/off conductance ratio. This effect
results from the inter-subband transmissions the probabilities of which are presented in
Figs.~\ref{fig11}(b) and (c). Both these figures clearly indicate that the reduction of $G(0)$ is
due to the inter-subband transmission with spin-flip. However, we note that $\alpha _{12}$ at
$V_g=0$ is nearly constant and is almost independent on
on the barrier width [Fig.~\ref{fig10}(c)]. Therefore, we conclude that the increase of
$T_{12}^{\uparrow \downarrow}$ is mainly caused by the reduction of $\Delta \varepsilon$ [cf.
Fig.~\ref{fig10}(d)], which decreases with the increasing barrier width - the reduction of the
coupling between quantum wells considerably weakens the repulsion of the states.	

\section{Summary}
\label{sec4}
The inter-subband-induced SO interaction in the quantum well with the double occupancy has attracted
the growing interest because it can give raise to interesting physical effects, e.g. unusual
Zitterbewegung. This specific SO interaction is nonzero
even in the symmetric heterostructure, as it arises from the coupling between states with opposite
parity. The strength of this coupling is
comparable to the ordinary Rashba intra-subband SO coupling. In the present paper we have analyzed
the influence of the inter-subband SO interaction on the spin transistor operation. For this
purpose, we have calculated the electron transport in the spin transistor within the two-subband
model including both the intra- and inter-subband SO interaction. We have started from the
model in which the SO coupling constants are treated as the parameters. In the absence of the
inter-subband SO interaction and with equal intra-subband SO coupling constants we have obtained the
regular conductance oscillations, similar to those predicted for the quantum well with the single
occupancy. We have shown that these oscillations are strongly affected by the inter-subband SO
interaction leading to its irregular and damped form. For large $\alpha _{12}$ we have found the
inversion of the oscillations, i.e., the conductance
maxima and minima interchange. Interestingly, we have demonstrated that even for the zero
intra-subband SO coupling related to the symmetric geometry, the conductance oscillates as a
function of the inter-subband SO coupling. This effect has been explained as resulting from the
inter-subband transitions with spin flip.  Finally we have also performed calculations with the
asymmetry of the intra-subband SO coupling constants. As we found the inter-subband SO interaction
lifts the symmetry of the conductance with respect to the subbands interchange.

In the second part of the paper, we have studied the conductance within the realistic spin
transistor
model with the conduction channel based on the Al$_{0.48}$In$_{0.52}$As/Ga$_{0.47}$In$_{0.53}$As
double quantum well. For the considered nanostructure, by performing a detailed self-consistent
calculations in which we solve both Poisson’s and Schr\"{o}dinger’s equation iteratively, we have
determined the strengths of the SO coupling contacts $\alpha _{11}$, $\alpha _{22}$ and
$\alpha_{12}$. The values of these coupling constants contain contributions arising from the
potential-well and barrier offsets, the Hartree potential, the external gate potential and the
modulation doping potential. We have obtained the resonant behavior of $\alpha
_{12}$ versus the gate voltage. Furthermore, the intra-subband SO coupling rapidly changes its sign
and magnitude at $V_g=0$. As we have stated in the paper such a rapid change of the SO coupling
constants in the narrow voltage range is favorable for the spin-FET application in which the
on/off conductance switching should be realized in the possibly narrow gate voltage. Our
calculations for different electron densities have shown that this effect can be strengthened for
the high electron
concentration in the quantum well. However, for the high electron density the inter-suband SO
interaction becomes dominant. The suppression of the conductance at $V_g=0$ which results from the
inter-subband transition with spin flip is the strong evidence of this interaction. This effect
leads to the reduction of the  on/off conductance ratio. Similar effect has been observed for the
wide central barrier, for which the increase of the inter-subband transmissions is mainly due to the
decrease of the energy separation between both the subbands with almost constant $\alpha _{12}$.

In summary, our studies of the inter-subband SO interaction on the spin transistor operation show
that this SO coupling leads to the reduction of the on/off conductance ratio and thus
decreases the efficiency of the spin transistor.

\begin{acknowledgements}
This work was supported by the funds of Ministry of Science and Higher Education for 2016 and  by
PL-Grid Infrastructure.
\end{acknowledgements}


%

\end{document}